\documentclass[10 pt, conference]{IEEEtran}
\IEEEoverridecommandlockouts                       
\usepackage{url}
\usepackage[utf8]{inputenc}
\usepackage[T1]{fontenc}
\usepackage{amsmath,amssymb,amsfonts}
\usepackage{graphicx}
\usepackage{subfigure}
\usepackage{textcomp}
\usepackage{xcolor}
\usepackage{verbatim}
\usepackage{makecell}
\usepackage{booktabs}
\usepackage{cite}
\usepackage{caption}
\usepackage{extarrows}
\usepackage{bm}
\usepackage{lettrine}
\usepackage[implicit=false]{hyperref}
\usepackage{amsthm}
\usepackage{algorithm,algorithmic}
\begin{document}
\title{Near-Field Beamforming Design for Multi-IRS Wireless Beam Routing} 
\author{\large{Tao Wang},
\large{Changsheng You}, \IEEEmembership{\large{Member IEEE}}, 
and \large{Changchuan Yin}, 
\IEEEmembership{\large{Senior Member IEEE}}
\thanks{This work was supported in part by Beijing Natural Science Foundation under Grant L223027, the National Natural Science Foundation of China under Grants 62471056 and 62201242, the 111 Project under Grant B17007, and the Program under Grant 2023QN10X152. 
\textit{(Corresponding author: Changchuan Yin.)}}
\thanks{T. Wang and C. Yin are with Beijing Laboratory of Advanced Information Networks, Beijing University of Posts and Telecommunications~(BUPT), Beijing 100876, China (e-mail: taowang@bupt.edu.cn, ccyin@bupt.edu.cn).}
\thanks{C. You is with the Department of Electrical and Electronic Engineering, Southern University of Science and Technology~(SUSTech), Shenzhen 518005, China (e-mail: youcs@sustech.edu.cn).}
}
\maketitle
\thispagestyle{empty}
\pagestyle{empty}

\begin{abstract}
This letter investigates efficient near-field beamforming designs for wireless communication systems aided by multiple extremely large-scale intelligent reflecting surfaces (Multi-XL-IRSs). Previous studies have mostly assumed far-field channel conditions, based on which angle-based beamforming is adopted for the base station (BS) and IRSs. However, this approach may lead to degraded performance in practical near-field scenarios. To address this issue, we consider near-field channel modeling and formulate an optimization problem aimed at maximizing the received power at the user equipment (UE) by jointly optimizing the beamforming at the BS and IRSs. Given the intractability under the beam training (BT)-based communication framework, the problem is revisited and reformulated. Subsequently, an efficient algorithm based on alternating optimization (AO) is proposed to solve the reformulated problem. Numerical results demonstrate that, in typical setups, our AO-based beamforming designs provide over 170\% and 20\% received power gains over conventional angle-based beamforming and beam focusing, respectively. Furthermore, the power scaling law derived under far-field conditions no longer holds in near-field scenarios.
\end{abstract}
\section{Introduction}

Intelligent reflecting surfaces (IRSs) have emerged as a transformative technology for wireless communication systems \cite{c8,you2024generationadvancedtransceivertechnologies,c4}. Comprising massive reconfigurable elements, IRSs dynamically adjust the signal phase/amplitude in a cost-effective, full-duplex manner. By strategically deploying IRSs on environmental surfaces, propagation environments can be programmatically controlled.
Extensive research has demonstrated IRSs' potential to enhance coverage in obstructed scenarios through multi-hop reflections over multiple IRSs, establishing virtual line-of-sight (LoS) paths between the base station (BS) and the user equipment (UE) \cite{single2multi}. This capability has spurred investigations into multi-IRS beam routing, which achieves superior cooperative passive beamforming (CPB) gains compared to single-IRS systems \cite{cooperative2021,multi-beam-multi-hop2022,MAMP2023}.

To mitigate multi-hop pathloss, it is necessary to deploy XL-IRSs with massive passive reflecting elements to achieve sufficient power capture and beamforming gains \cite{Power-Scaling-Laws-and-Near-Field-Behaviors}. 
Notably, inter-array channels in such deployments often exhibit near-field characteristics because the near-far field boundary is proportional to the array size \cite{MIMO-RD-ARD-distance}\footnote{This work focuses on the near-field condition caused by passive XL-IRSs. Active IRSs can amplify reflected signals, thus reducing the required number of active elements \cite{Active_STAR_RIS}. Therefore, active IRSs are typically manifested as small-scale arrays and adhere to far-field channel models.}. However, existing studies are predominantly based on simplified far-field channel models \cite{single2multi,cooperative2021,multi-beam-multi-hop2022, MAMP2023}, leading to angle-based beamforming designs, which can result in degraded performance in practical near-field conditions due to channel mismatch.
Although our previous work \cite{BSbeamforming} addressed the near-field beamforming design in single-IRS systems, it cannot be applied to multi-IRS systems.

Motivated by the above, this letter proposes a joint beamforming design scheme for BS and IRSs under near-field channel modeling to optimize multi-IRS beam routing. 
We formulate a received power maximization problem under beam training (BT) protocols, which proves intractable. To address this issue, we reformulate the original problem and propose an alternating optimization (AO)-based algorithm to solve the reformulated problem. 
Numerical results depict the power scaling of the multi-IRS beam routing systems and demonstrate the superiority of the proposed AO-based beamforming design over the benchmark schemes.

\textit{Notations}: 
Variables, column vectors, and matrices are denoted by normal-face letters (e.g., $x$), bold-face lower-case letters (e.g., $\mathbf{x}$), and bold-face upper-case letters (e.g., $\mathbf{X}$), respectively. 
$[\mathbf{x}]_n$ denotes the $n$-th element of a vector. $[\mathbf{X}]_{n,m}$ denotes the $(n,m)$-th element of a matrix.
$(\cdot)^T$ and $(\cdot)^H$ denote the transpose and conjugate transpose, respectively. $\textrm{dist}(\mathbf{x},\mathbf{y}) = \|\mathbf{x} - \mathbf{y}\|_2$ calculates the Euclidean distance between two coordinates. $\angle \mathbf{X}$ normalizes the amplitude of each element in a matrix (phase-only preservation). $\textrm{diag}(\mathbf{x})$ creates a diagonal matrix from a vector. $\textrm{diagvec}(\mathbf{X})$ extracts diagonal elements of a matrix into a column vector. $\Vert\cdot\Vert_1$ and $\Vert\cdot\Vert_2$ denote the $\ell_1$-norm and $\ell_2$-norm, respectively. Lastly, $\mathcal{CN}(0, \sigma^2)$ represents the circularly symmetric complex Gaussian distribution.

\section{System Model}
\begin{figure}[t]
	\centering
	\includegraphics[scale = 0.13]{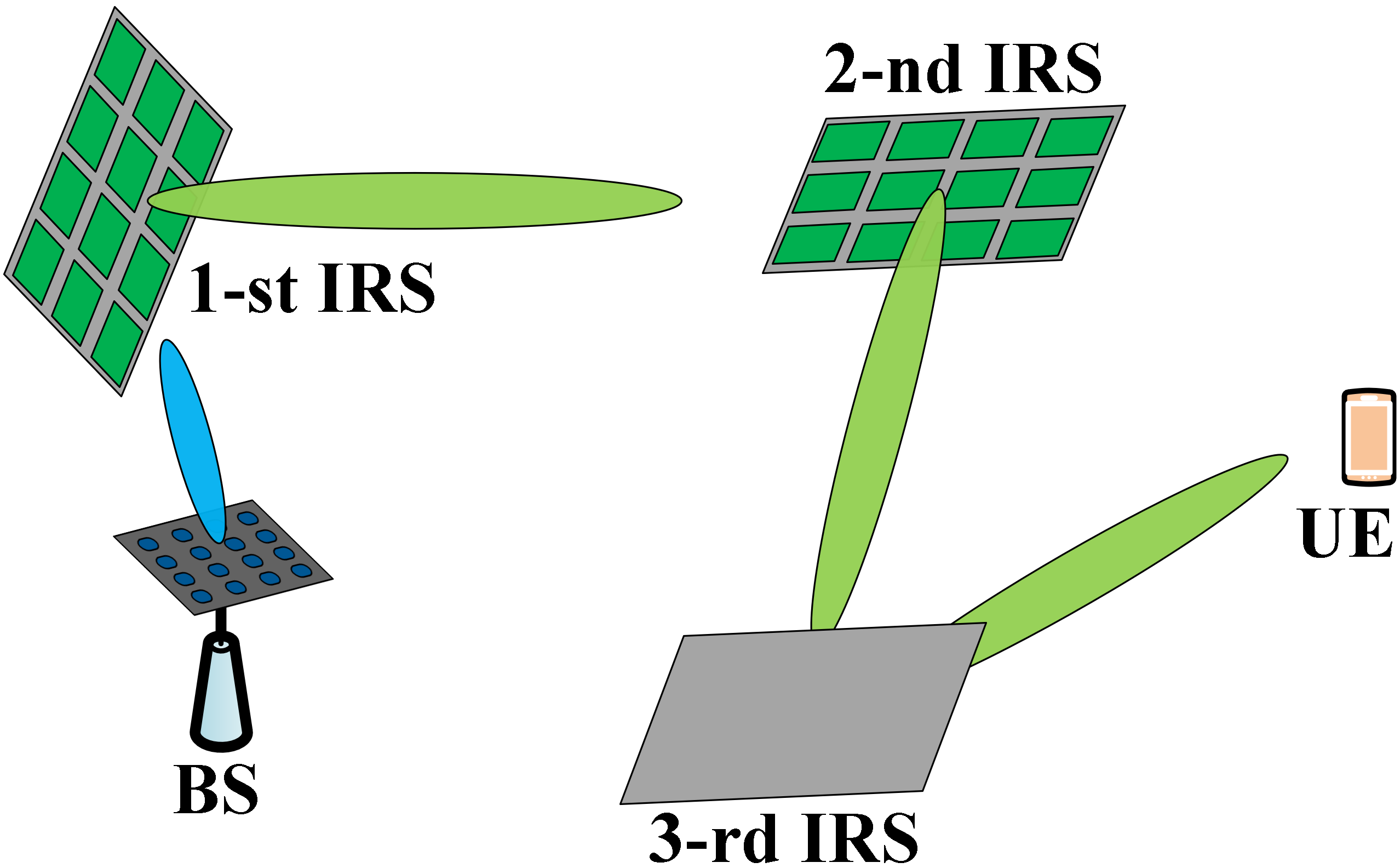}
    \caption{A Multi-IRS-assisted wireless beam routing system.}
   \label{fig:system}
\end{figure}

As depicted in Fig. \ref{fig:system}, we consider a multi-IRS-assisted beam routing system, where the direct BS-UE link is obstructed, while a virtual LoS link is established via multi-hop reflections over $K \in \mathbb{N}^+$ IRSs. The BS is equipped with $M$ antennas, indexed by $\mathcal{M} = \{1, \cdots, M\}$. The $k$-th IRS is composed of $N_k$ elements, indexed by $\mathcal{K} = \{1, 2, \cdots, k, \cdots, K\}$ and $\mathcal{N}_k = \{1, 2, \cdots, N_k\}$. For simplicity, the index $k$ also equals the hop count from the BS to the $k$-th IRS. 
A single-antenna UE is within the coverage area of the $K$-th IRS\footnote{If there are multiple UEs within the coverage area of the $K$-th IRS, the results of the proposed algorithm can be directly applied to IRS beam training for these multiple UEs.}. 
The control links from the BS to the IRSs and the feedback link from the UE to the BS are assumed to be separate and reliable.

To model the near-field LoS channels, the coordinates of the $m$-th BS antenna, $n_k$-th IRS element, and UE are denoted as $\mathbf{c}_m^{\rm{B}}, \mathbf{c}_{n_k}^{\rm{I}}$, and $\mathbf{c}^{\rm{U}}$, respectively, where $m \in \mathcal{M}, n_k \in \mathcal{N}_k, k \in \mathcal{K}$. Consequently, the distance from the $m$-th BS antenna to the $n_1$-th IRS element is given by $d_{n_1,m} = \textrm{dist}(\mathbf{c}_{n_1}^{\rm{I}},\mathbf{c}_m^{\rm{B}})$. The distance from the $n_{k-1}$-th IRS element to the $n_{k}$-th IRS element is $d_{n_{k},n_{k-1}} = \textrm{dist}(\mathbf{c}_{n_{k}}^{\rm{I}},\mathbf{c}_{n_{k-1}}^{\rm{I}})$, where $2\leq k \leq K$. The distance from the $n_K$-th IRS element to the UE is $d_{n_K} = \textrm{dist}(\mathbf{c}_{n_K}^{\rm{I}}, \mathbf{c}^{\rm{U}})$.

\subsection{Signal Model}
Let $\mathbf{H}_1 \in \mathbb{C}^{N_1 \times M}$ represent the channel from the BS to the first IRS, $\mathbf{H}_{k} \in \mathbb{C}^{N_{k}\times N_{k-1}},~2\le k\le K$ represent the channel from the ($k-1$)-th IRS to the $k$-th IRS, and $\mathbf{h}^H \in \mathbb{C}^{1\times N_K}$ represent the channel from the $K$-th IRS to the UE. 
For the wireless beam routing system, we focus on the cascaded LoS link by treating the non-LoS (NLoS) components as part of the environment scattering, which has been shown to have a negligible impact on system performance \cite{cooperative2021, multi-beam-multi-hop2022}.
Therefore, considering only the LoS component, the near-field channel between the BS and the first IRS is modeled as \cite{liuYW2023NFC_Tutorial}
\begin{equation}   
 \label{BS-IRS_channel}
 \begin{split}
 [\mathbf{H}_1]_{n_1,m} = \sqrt{\frac{G^{\rm B}_{n_1,m}
 A_{n_1,m}^{\rm I}}
 {4\pi d^2_{n_1,m}}}e^{j\frac{2\pi}{\lambda}d_{n_1,m}},~n_1 \in \mathcal{N}_1, m\in \mathcal{M}, 
 \end{split}
 \end{equation}
where $G^{\rm B}_{n_1,m}$ is the antenna gain from the $m$-th BS antenna to the $n_1$-th IRS element, and $A_{n_1,m}^{\rm I}$ is the effective aperture of the $n_1$-th IRS element relative to the $m$-th BS antenna. Similarly, we have
\begin{equation}   
 \label{inter-IRS_channel}
 \begin{split}
 [\mathbf{H}_k]_{n_{k},n_{k-1}} &= \sqrt{\frac{G^{\rm I}_{n_{k},n_{k-1}}
 A_{n_{k},n_{k-1}}^{\rm I}}
 {4\pi d^2_{n_{k},n_{k-1}}}}e^{j\frac{2\pi}{\lambda}d_{n_{k},n_{k-1}}}, 
 \end{split}
 \end{equation}
where $n_{k}\in\mathcal{N}_{k}$, $n_{k-1}\in\mathcal{N}_{k-1}$, $2\le k\le K$. $G^{\rm I}_{n_{k},n_{k-1}}$ is the element gain from the $n_{k-1}$-th IRS element to the $n_{k}$-th IRS element. $A_{n_{k},n_{k-1}}^{\rm I}$ is the effective aperture of the $n_{k}$-th IRS element relative to the $n_{k-1}$-th IRS element. Moreover,
\begin{equation}   
 \label{IRS-UE_channel}
 \begin{split}
 [\mathbf{h}^H]_{n_{K}} &= \sqrt{\frac{G^{\rm I}_{n_K}
 A^{\rm U}_{n_K}}
 {4\pi d^2_{n_K}}}e^{j\frac{2\pi}{\lambda}d_{n_K}}, 
 \end{split}
 \end{equation}
where $G^{\rm I}_{n_{K}}$ is the element gain from the $n_{K}$-th IRS element to the UE antenna. $A^{\rm U}_{n_{K}}$ is the effective aperture of the UE antenna relative to the $n_{K}$-th IRS element. 

Based on the above, the received signal at the UE is
\begin{equation}   
 \label{received_signal}
 \begin{split}
 y &= \mathbf{h}^H
 \mathbf{\Phi}_K
 \mathbf{H}_K\cdots\mathbf{\Phi}_1
 \mathbf{H}_1
 \mathbf{w}s+z,
 \end{split}
 \end{equation}
where $s$ is the transmitted symbol, $z \sim \mathcal{C}\mathcal{N}(0, {\sigma^2})$ is the additive white Gaussian noise (AWGN), $\mathbf{w} \in \mathbb{C}^{M\times 1}$ is the BS beamforming vector subject to the power constraint $\Vert\mathbf{w}\Vert_2^2 \leq P_{\rm t}$. $\mathbf{\Phi}_k={\rm{diag}}( e^{j\varphi_{k,1}}, \cdots, e^{j\varphi_{k, N_k}})\in \mathbb{C}^{N_k\times N_k}$ represents the phase shifts introduced by the $k$-th IRS, where $\varphi_{k, n_k} \in [0,2\pi), k\in\mathcal{K}, n_k\in\mathcal{N}_k$. Consequently, the received signal power at the UE is
\begin{equation}
 \label{received-power}
 \begin{aligned}
  P_{\rm r} &= \left|\mathbf{h}^H
 \mathbf{\Phi}_K
 \mathbf{H}_K\cdots\mathbf{\Phi}_1
 \mathbf{H}_1
 \mathbf{w}\right|^2.
  \end{aligned}
\end{equation}
The achievable data rate in bit/second/Hz is calculated by 
\begin{equation}
 \label{achievable-data-rate}
 \begin{aligned}
  R = \log_2(1+\frac{P_{\mathrm{r}}^2}{\sigma^2}).
  \end{aligned}
\end{equation}

Since the BS and IRSs are usually strategically deployed at fixed locations, we assume that the inter-array LoS channels, i.e., $\mathbf{H}_k, k\in \mathcal{K}$, are known \emph{a priori}, which can be obtained from their deployment geometry \cite{liuYW2023NFC_Tutorial}. In contrast, due to mobility of the UE, the channel $\mathbf{h}^H$ is random and unknown. 

\subsection{BT-Based Transmission Protocol}
\label{BT-Based Transmission Protocol}
For practical IRS implementations, codebook-based BT is usually preferred to establish an initial link, especially when the channel dimension is large or the signal-to-noise ratio (SNR) is low \cite{codebook-IRS-Dai}. Specifically, under the control of the BS, the $K$-th IRS performs beam sweeping based on a predefined codebook, and the UE measures the signal powers of these beams and feeds the index of the selected codeword to the BS. After BT, the BS applies the selected codeword at the $K$-th IRS to establish an initial link for subsequent channel estimation and data transmissions.
During the BT process, only the phase shifts of the $K$-th IRS are adjusted for beam sweeping, while the beamforming vectors of the BS and intermediate IRSs, i.e., $\mathbf{w}$ and $\mathbf{\Phi}_{k},~k=1,2,\cdots, K-1$, are pre-designed and held fixed \cite{Fast-Near-Field-Beam-Training-IRS, WCNC2024}. 
Therefore, the procedure of the BT-based transmission protocol can be summarized as follows. 
\begin{enumerate}
    \item First, $\mathbf{w}$ and $\mathbf{\Phi}_{k},~k=1,2,\cdots, K-1$, are optimized and pre-configured based on prior inter-array LoS channel information. 
    \item Second, BT is performed at the $K$-th IRS to determine the optimal codeword for the $K$-th IRS, i.e., $\mathbf{\Phi}_{K}$. 
    \item Third, with $\mathbf{w}$ and $\mathbf{\Phi}_{k},~k=1,2,\cdots, K$ all fixed, a low-dimensional channel estimation is performed for the equivalent baseband BS-UE channel, followed by data transmission. \end{enumerate}

\subsection{Problem Formulation}
Based on the BT-based communication protocol, this letter aims to develop an efficient joint beamforming design approach for the BS and intermediate IRSs, given the inter-array near-field LoS channels. To this end, a received power maximization problem is formulated as
\begin{equation}   
 \label{BS-beamforming-problem-1}
 \begin{split}
\max_{\mathbf{w},~\mathbf{\Phi}_{k},~k=1,2,\cdots,K-1}
~&\Vert\mathbf{h}^H
 \mathbf{\Phi}_K
 \mathbf{H}_K\cdots\mathbf{\Phi}_1
 \mathbf{H}_1
 \mathbf{w}\Vert_2^2,\\
 {\rm s.t.}~ &\Vert\mathbf{w}\Vert_2^2 \leq P_{\rm t},\\
 ~&\varphi_{k, n_k} \in [0,2\pi), k\in\mathcal{K}, n_k\in\mathcal{N}_k,\\
 ~&\mathbf{\Phi}_K \in \mathcal{C},
 \end{split}
 \tag{P1}
\end{equation}
where $\mathcal{C}$ represents the predefined codebook used for BT\footnote{The codebook-based BT has been widely studied in the literature \cite{multi-beam-training-IRS,Fast-Near-Field-Beam-Training-IRS, WCNC2024, 2D_DFT_DA}, so it is not studied in this letter.}. It is evident that the objective function in (\ref{BS-beamforming-problem-1}) reaches its maximum when $\Vert\mathbf{w}\Vert_2^2 = P_{\rm t}$. 
According to the BT-based transmission protocol, the optimization of $\mathbf{w}$ and $\mathbf{\Phi}_{k},~k=1,2,\cdots, K-1$ occurs before the BT at the $K$-th IRS, when both $\mathbf{h}^H$ and $\mathbf{\Phi}_K$ are unknown. This renders the optimization of $\mathbf{w}$ and $\mathbf{\Phi}_{k},~k=1,2,\cdots, K-1$ infeasible.
Consequently, (\ref{BS-beamforming-problem-1}) cannot be solved directly.

\subsection{Benchmark Schemes}
\subsubsection{Conventional Angle-based Beamforming Design}
To address (\ref{BS-beamforming-problem-1}), prior research \cite{single2multi, cooperative2021, multi-beam-multi-hop2022, MAMP2023} has mostly utilized angle-based beamforming for $\mathbf{w}$ and $\mathbf{\Phi}_{k},~k=1,2,\cdots, K-1$ assuming far-field channel modeling. In this approach, given the assumed inter-array far-field LoS channels, each array aligns its transmit/reflect beamforming phases with the angle of departure (AoD) directed toward the subsequent array.
However, for multi-IRS wireless beam routing systems, it is necessary to employ near-field channel modeling for accurate channel characterization \cite{liuYW2023NFC_Tutorial}, rendering the existing angle-based beamforming method inapplicable due to channel mismatch. 
\subsubsection{Near-field Beam Focusing}
We come up with an intuitive beamforming design of $\mathbf{w}$ and $\mathbf{\Phi}_{k},~k=1,2,\cdots, K-1$, which is named near-field beam focusing. In this approach, given the inter-array near-field LoS channels, each array aligns its transmit/reflect beamforming phases at the center point of the subsequent array. According to the diffraction limit, the signal energy is distributed mainly in the region centered around the focal point \cite{c9}. Thus, intuitively, the total received power tends to be maximized in each array.

\section{Proposed AO-Based Beamforming Design}
An efficient AO-based algorithm is proposed to resolve (\ref{BS-beamforming-problem-1}). Specifically, we adopt the uniform amplitude approximation for $\mathbf{h}^H$, which is based on the Fresnel approximation and can adequately address the near-field multiple-input single-output (MISO) channels, as concluded in \cite{liuYW2023NFC_Tutorial, Power-Scaling-Laws-and-Near-Field-Behaviors}. Consequently, we have $[\mathbf{h}^H\mathbf{\Phi}_K]_{n_{K}} = \sqrt{\frac{G^{\rm I} A^{\rm U}} {4\pi d^2_{n_K}}}e^{j(\frac{2\pi d_{n_K}}{\lambda}+\varphi_{K, n_K})}\approx \alpha_K e^{j(\frac{2\pi d_{n_K}}{\lambda}+\varphi_{K, n_K})}$, where $\alpha_K$ represents the uniform amplitude of $\mathbf{h}^H$. Given that $\varphi_{K, n_K}, n_K \in \mathcal{N}_K$ are adjustable within $[0, 2\pi)$, we introduce an auxiliary variable, $\boldsymbol{\psi}^{T} \in \mathbb{C}^{1\times N_K}$, and set $\alpha_K \boldsymbol{\psi}^{T} \equiv \mathbf{h}^H \mathbf{\Phi}_K$, where $\boldsymbol{\psi}^T$ is restricted by $\left|[ \boldsymbol{\psi}^{T}]_{n_K}\right| = 1, \forall~n_K\in \mathcal{N}_K$. With this transformation, (\ref{BS-beamforming-problem-1}) is transformed into 
\begin{equation}   
 \label{BS-beamforming-problem-2}
 \begin{split}
\max_{\mathbf{w},\boldsymbol{\psi}^{T},\mathbf{\Phi}_{k},~k=1,2,\cdots,K-1}&
~\Vert \boldsymbol{\psi}^T
 \mathbf{H}_K\cdots\mathbf{\Phi}_1
 \mathbf{H}_1
 \mathbf{w}\Vert_2^2,\\
 {\rm s.t.}~ &\Vert\mathbf{w}\Vert_2^2 = P_{\rm t},\\
  ~&\varphi_{k, n_k} \in [0,2\pi), k\in\mathcal{K}, n_k\in\mathcal{N}_k,\\
~&\left|[\boldsymbol{\psi}^{T}]_{n_K}\right| = 1,~n_K\in\mathcal{N}_K,
 \end{split}
 \tag{P2}
\end{equation}
where the intractability caused by the BT-based transmission protocol in (\ref{BS-beamforming-problem-1}) is eliminated. However, (\ref{BS-beamforming-problem-2}) is still non-convex due to its non-convex objective and unit-modulus constraints. Moreover, high-dimensional variables render exhaustive or heuristic algorithms infeasible because of their prohibitively high complexity. To address this problem, we propose an efficient AO-based algorithm that determines one variable at a time while keeping the others fixed in each iteration, as shown in Algorithm 1.

For the update of $\mathbf{w}$, since ${\boldsymbol{\psi}^{T}} \mathbf{H}_K\cdots\mathbf{\Phi}_1 \mathbf{H}_1 \in \mathbb{C}^{1 \times M}$ and $\mathbf{w} \in \mathbb{C}^{M \times 1}$, based on the Cauchy-Schwarz inequality, the update of $\mathbf{w}$ should be $\mathbf{\widetilde{w}} \leftarrow \left ({\boldsymbol{\psi}^{T}} \mathbf{H}_K\cdots\mathbf{\Phi}_1 \mathbf{H}_1\right )^H$, plus a normalization represented by $\mathbf{w} \leftarrow \sqrt{P_{\rm t}} \mathbf{\widetilde{w}}/ \Vert \mathbf{\widetilde{w}}\Vert_2$.

For the update of $\mathbf{\Phi}_{k},~k =1,2,\cdots, K-1$, let us take $\mathbf{\Phi}_1$ as an example. Since $\boldsymbol{\psi}^{T}\mathbf{H}_K \cdots \mathbf{\Phi}_{2}\mathbf{H}_2 \in \mathbb{C}^{1 \times N_1}$ and $\mathbf{H}_1\mathbf{w} \in \mathbb{C}^{N_1 \times 1}$, we have $\boldsymbol{\psi}^T
 \mathbf{H}_K\cdots\mathbf{\Phi}_1
 \mathbf{H}_1
 \mathbf{w} = (\boldsymbol{\psi}^{T}\mathbf{H}_K \cdots \mathbf{\Phi}_{2}\mathbf{H}_2) \odot (\mathbf{H}_1\mathbf{w})^T \cdot \textrm{diagvec}(\mathbf{\Phi}_1)$. With this transformation, the update of $\mathbf{\Phi}_1$ should follow $\textrm{diagvec}(\mathbf{\Phi}_1) = \angle\left((\boldsymbol{\psi}^{T}\mathbf{H}_K \cdots \mathbf{\Phi}_{2}\mathbf{H}_2) \odot (\mathbf{H}_1\mathbf{w})^T\right)^H$ to align the phases of $(\boldsymbol{\psi}^{T}\mathbf{H}_K \cdots \mathbf{\Phi}_{2} \mathbf{H}_2) \odot (\mathbf{H}_1\mathbf{w})^T$, which is
$\mathbf{\Phi}_1 \leftarrow \mathrm{diag}\left(\angle\left((\boldsymbol{\psi}^{T}\mathbf{H}_K  \cdots \mathbf{\Phi}_{2} \mathbf{H}_2) \odot (\mathbf{H}_1\mathbf{w})^T\right)^H\right)$. Similar derivations can be applied to obtain the updates for $\mathbf{\Phi}_{k},~k = 2,\cdots, K-1$, which are shown in Steps 10 and 13 of Algorithm 1. 

For the update of $\boldsymbol{\psi}^{T}$, $\boldsymbol{\psi}^{T} \leftarrow \angle \left(\mathbf{H}_K\cdots\mathbf{\Phi}_1 \mathbf{H}_1 \mathbf{w}\right )^{H}$ is performed to achieve phase alignment of $\mathbf{H}_K\cdots\mathbf{\Phi}_1 \mathbf{H}_1 \mathbf{w}$. 

The above analysis implies that every update in Algorithm 1 is non-decreasing. Therefore, Algorithm 1 is guaranteed to converge monotonically to a local stationary point. However, convergence to the global optimal point cannot be guaranteed due to the non-convexity of the original problem.

\floatname{algorithm}{Algorithm}
\begin{algorithm}[t]
\label{algorithm-1}
	\caption{AO-based beamforming design}
	\begin{algorithmic}[1]
\STATE \textbf{Input}: $\mathbf{H}_k, k\in\mathcal{K}$, number of iterations $I_\mathrm{max}$.
\STATE \textbf{Initialization}: Initialize $\boldsymbol{\psi}^{T}$, and $\mathbf{\Phi}_{k},~k =1,2,\cdots, K-1$, which satisfies unit-modulus constraint for each of their elements. 
\FOR{ $i = 1,2, \cdots, I_\mathrm{max}$}
\STATE \textbf{ // Design the BS beamforming.}
\STATE $\mathbf{\widetilde{w}} \leftarrow \left ({\boldsymbol{\psi}^{T}} \mathbf{H}_K\cdots\mathbf{\Phi}_1 \mathbf{H}_1\right )^H$
\STATE $\mathbf{w} \leftarrow \sqrt{P_{\rm t}} \mathbf{\widetilde{w}}/ \Vert \mathbf{\widetilde{w}}\Vert_2$
\STATE \textbf{ // Design the IRS beamforming.}
\STATE $\mathbf{\Phi}_1 \leftarrow \mathrm{diag}\left(\angle\left((\boldsymbol{\psi}^{T}\mathbf{H}_K  \cdots \mathbf{\Phi}_{2}\mathbf{H}_2) \odot (\mathbf{H}_1\mathbf{w})^T\right)^H\right)$
\FOR{$k = 2, \cdots, K-2$}
\STATE $\mathbf{\Phi}_k  \leftarrow \mathrm{diag}\Big{(}\angle\big{(}(\boldsymbol{\psi}^{T}\mathbf{H}_K\cdots\mathbf{\Phi}_{k+1}\mathbf{H}_{k+1})~\odot$
\STATE $~~~~~~~~(\mathbf{H}_{k}\mathbf{\Phi}_{k-1}\cdots\mathbf{H}_1\mathbf{w})^T\big{)}^H\Big{)}$
\ENDFOR
\STATE $\mathbf{\Phi}_{K-1} \leftarrow \mathrm{diag}\Big{(}\angle\big{(}(\boldsymbol{\psi}^{T}\mathbf{H}_K)~\odot$ 
\STATE $~~~~~~~~~~~(\mathbf{H}_{K-1}\mathbf{\Phi}_{K-2}\cdots\mathbf{H}_1\mathbf{w})^T\big{)}^H\Big{)}$
\STATE \textbf{ // Determine the auxiliary variable.}
\STATE $\boldsymbol{\psi}^{T} \leftarrow \angle \left(\mathbf{H}_K\cdots\mathbf{\Phi}_1 \mathbf{H}_1 \mathbf{w}\right )^{H}$
\ENDFOR
\STATE \textbf{Output}: $\mathbf{w}$, and $\mathbf{\Phi}_{k},~k =1,2,\cdots, K-1$.
\end{algorithmic}
\end{algorithm}        

\subsection{Complexity Analysis}
Without loss of generality, the number of elements in each IRS is uniformly denoted as $N$.
For the proposed Algorithm 1, the complexity is determined by the update of $\mathbf{w}, \mathbf{\Phi}_{k},~k =1,2,\cdots, K-1$, $\boldsymbol{\psi}^{T}$ in each iteration, and the number of iterations $I_{\mathrm{max}}$, which is approximately $\mathcal{O}\left(I_{\mathrm{max}}K^2N^2\right)$. Since the inter-array LoS channels are assumed to be quasi-static, ideally the Algorithm 1 can be performed only once in an offline manner.

\section{Numerical Results}
\label{Numerical Results}
In this section, Monte Carlo simulations are conducted to evaluate the proposed \emph{AO-based beamforming}. Two benchmarks are adopted: 1) \emph{angle-based beamforming}; 2) \emph{beam focusing}. For BT at the $K$-th IRS, the scheme proposed in \cite{2D_DFT_DA}, which involves a two-dimensional (2D) angle-domain BT followed by a one-dimensional (1D) distance-domain BT, is adopted.
 Unless otherwise specified, uniform rectangular arrays (URA) with half-wavelength spacing are considered for BS and IRSs, which are placed parallel and equidistant \cite{liuYW2023NFC_Tutorial}\footnote{The proposed scheme is independent of the channel dimensions, allowing the analytical results to be readily extended to other array geometries.}. The inter-array distance is 5 m, and the UE distance (from the center of the $K$-th IRS to the UE) is 20 m. The other parameters are listed in Table \ref{parameters}. To simulate the unknown NLoS components, the Rician factor is set to 20 dB. 
Based on the transmission protocols in Section \ref{BT-Based Transmission Protocol}, the inter-array LoS channels are used to obtain the beamforming of the BS and intermediate IRSs, while channels including both LoS and NLoS components are used during BT. 
The resulting points are averaged over 100 channel realizations.

 \begin{table}[t]
	\centering  
	\caption{Parameter configuration}  
	\label{parameters}  
	\begin{tabular}{cc|cc}  
		        \hline  
		& & & \\[-6pt]  
		Parameters&Values&Parameters&Values \\  
		        \hline
		& \\[-6pt]  
		$P_{\rm t}$&20 dBm & $\sigma^2$ & -70~dBm\\
		        \hline
    	& \\[-6pt]
        $M$&256&$N_k,~k \in \mathcal{K}$&4096 \\
		        \hline
    	& \\[-6pt]
		$K$ & 3 & $\lambda$ & 0.02~m \\
		        \hline
        & \\[-6pt]  
		$G^{\rm B},G^{\rm I}$&1&$A^{\rm U}$&$\lambda^2/(4\pi)$ \\
		        \hline
	\end{tabular}
\end{table}

\begin{figure}[t]
    \centering
    \subfigure[Angle-based beamforming.]
    {
        \includegraphics[scale = 0.268]{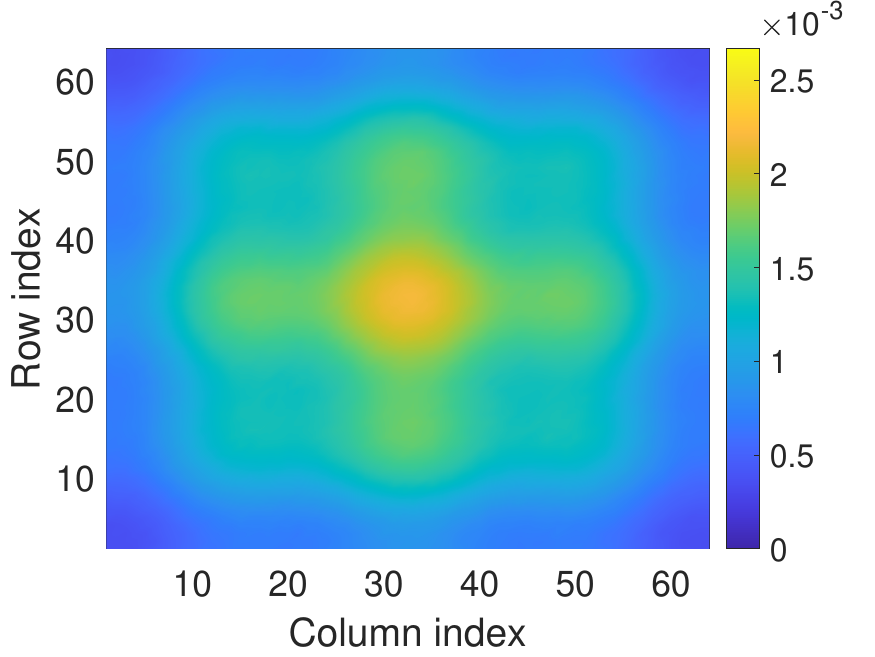}
	\label{fig:angle-based beamforming}
    }
    \subfigure[Beam focusing.]
    {
        \includegraphics[scale = 0.268]{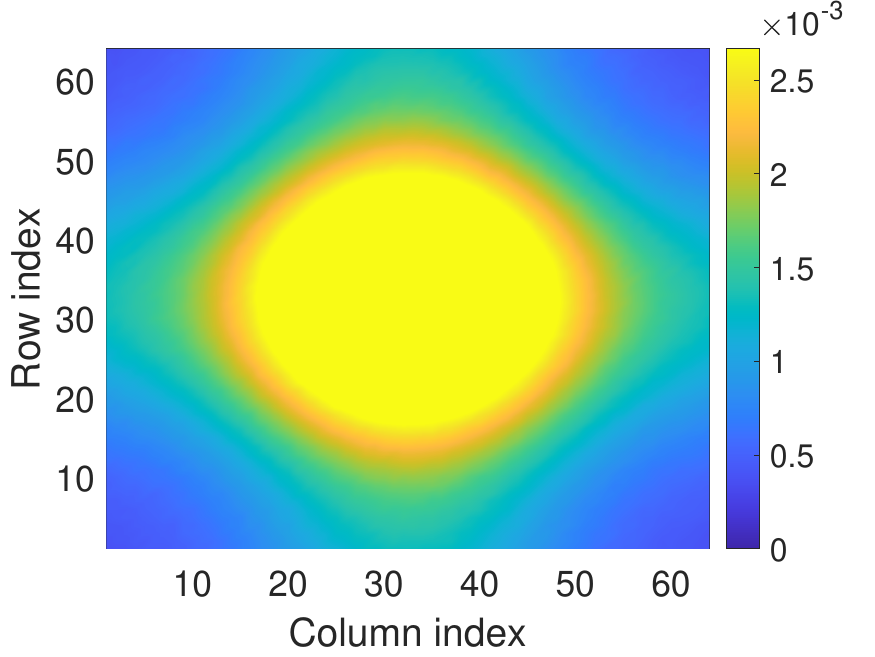}
	\label{fig:beam focusing}
    }
\quad    
    \subfigure[AO-based beamforming.]
    {
        \includegraphics[scale = 0.268]{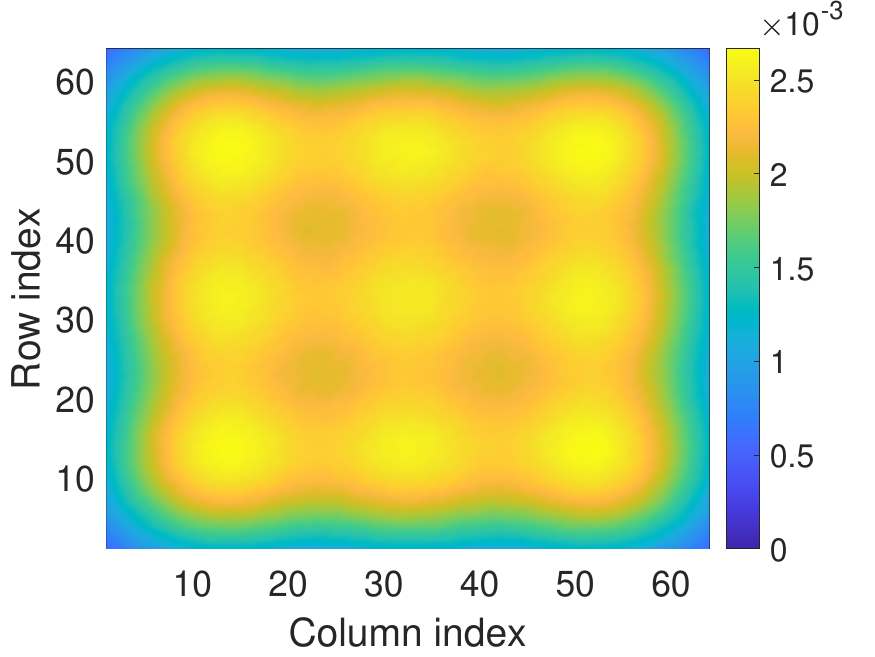}
	\label{fig:AO-256-4096}
    }
\caption{Incident signal amplitudes on the $K$-th IRS for different schemes, where the total incident signal powers are 0.0077, 0.0128, 0.0134 Watts, respectively.}
    \label{Signal-Amplitudes}
\end{figure}

Fig. \ref{Signal-Amplitudes} shows the incident signal amplitudes on the $K$-th IRS.
The total power of the incident signal under angle-based beamforming is very low due to channel mismatch. The incident signal under beam focusing is largely confined within the central region of the array, which aligns with the diffraction limit.
Compared with angle-based beamforming and beam focusing, the proposed AO-based beamforming achieves a larger total incident signal power and a more uniform distribution (meaning greater array gain) of the incident signal power. 
These observations demonstrate that the AO-based scheme can effectively improve both the total incident power and the array gain of the $K$-th IRS.

\begin{figure}[t]
    \centering
    \subfigure[Convergence of Algorithm 1.]
    {\includegraphics[scale = 0.285]{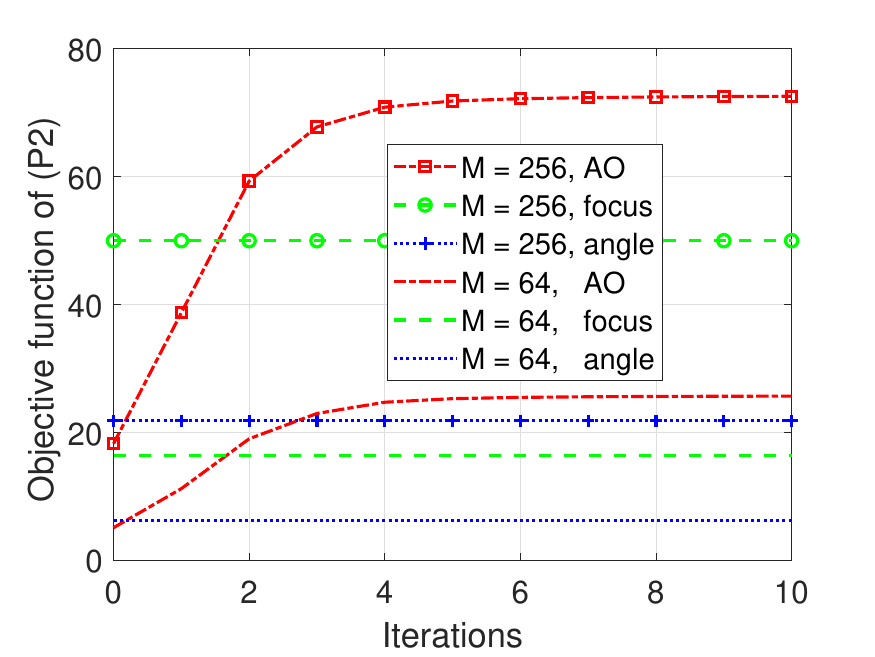}
	\label{fig:convergence}}     
     \subfigure[Received signal power v.s. $N$.]
    {\includegraphics[scale = 0.285]{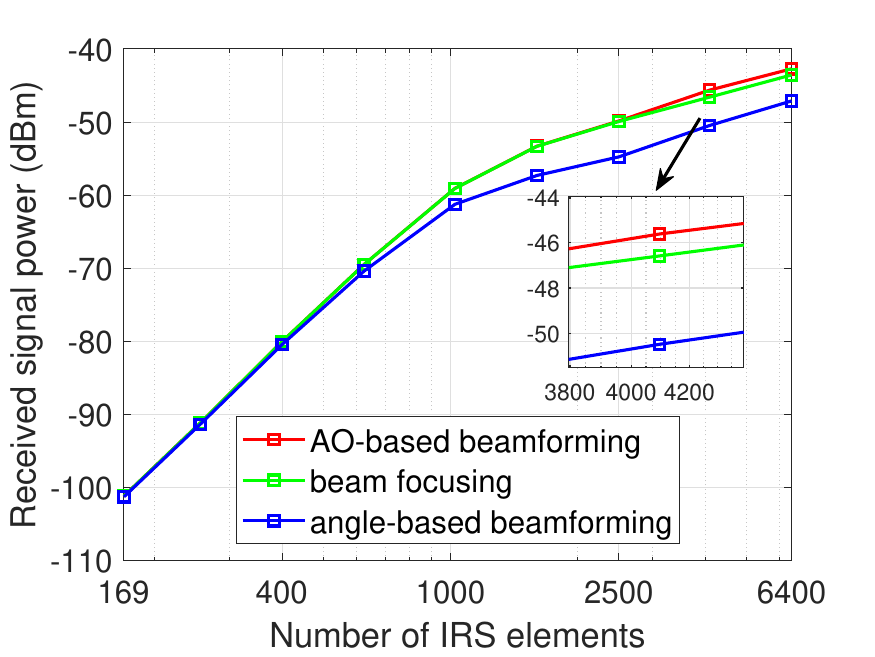}	
	\label{fig:N-IRS}}
\quad    
    \subfigure[Data rate v.s. UE distance.]
    {\includegraphics[scale = 0.285]{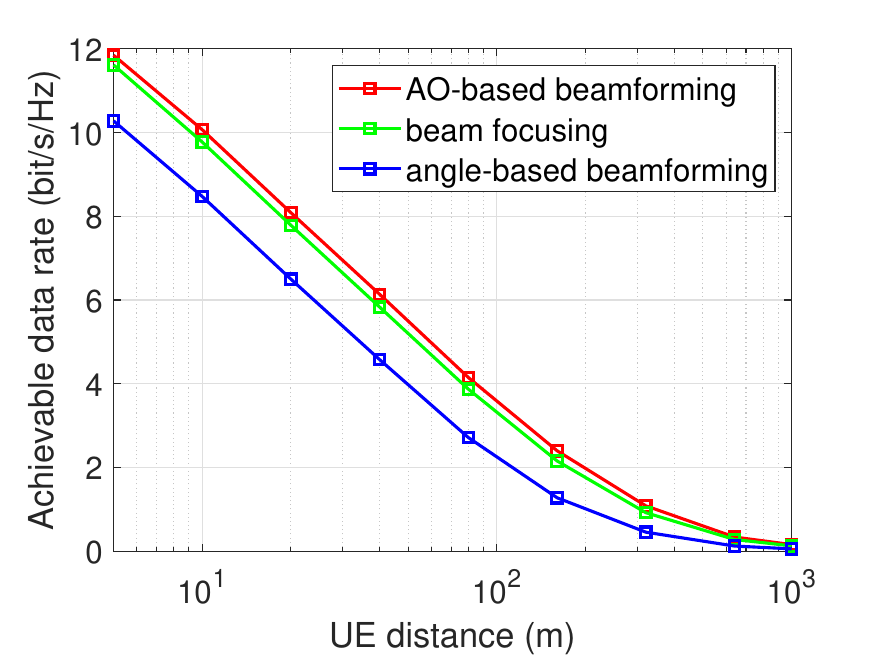}
    \label{fig:UE distance}}   
    \subfigure[Data rate v.s. Rician factor.]
    {\includegraphics[scale = 0.285]{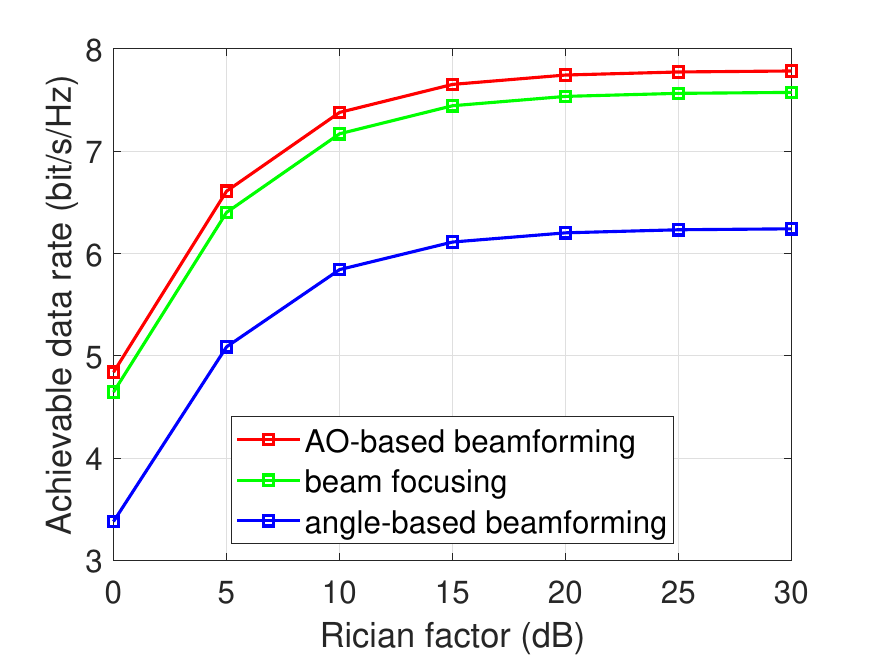}
	\label{fig:Rician factor}}
\caption{Convergence of Algorithm 1, and performance comparison of the schemes.}
    \label{simulation-results}
\end{figure}

Fig. \ref{fig:convergence} shows the convergence of Algorithm 1. It is shown that Algorithm 1 converges within 6 iterations for various parameters. Furthermore, the convergence values of the proposed AO-based scheme are significantly higher than those of the benchmark schemes, showing its effectiveness in solving (\ref{BS-beamforming-problem-2}).

Fig. \ref{fig:N-IRS} shows the impact of the number of IRS elements on the received signal power at the UE. 
The number of elements in each IRS is uniformly denoted as $N$, which varies from 169 to 6400, which means that the side length of each IRS varies from 0.13 to 0.8 m. 
Since 3 IRSs are deployed, the CPB gain derived under far-field assumptions is $\mathcal{O}(N^6)$ according to \cite{cooperative2021}, which means that $P_{\rm{r}}$ and $N$ exhibit a log-linear relationship. This is verified in Fig. \ref{fig:N-IRS} when $N$ is approximately less than 400, where far-field assumptions hold. In contrast, as $N$ continues to increase, the logarithmic linear power scaling no longer holds due to the failure of the far-field assumptions. Furthermore, as $N$ approaches infinity, the received signal power will saturate according to the principle of energy conservation \cite{Power-Scaling-Laws-and-Near-Field-Behaviors}.

Furthermore, in the large $N$ regime of Fig. \ref{fig:N-IRS}, the AO-based scheme achieves significant gains, e.g., more than 170\% and 20\% at $N = 4096$, over those of the angle-based beamforming and beam focusing, respectively. In the small $N$ regime, where far-field assumptions hold, the AO-based scheme performs comparably to the benchmark schemes, indicating that the proposed scheme is applicable in both near-field and far-field regions. 
Besides, we notice that the received signal power is too low to support viable data transmission in the small $N$ regime, e.g., less than -70 dBm when $N<625$. Therefore, multi-IRS beam routing systems should practically operate in the large $N$ regime, where the proposed AO-based scheme can be applied to improve system performance significantly.

Fig. \ref{fig:UE distance} shows the impact of the UE distance on the achievable data rate. The AO-based scheme outperforms the benchmark schemes across different UE distances, since higher received signal power is achieved, as shown in Fig. \ref{fig:N-IRS}. For the same reason, the performance of the proposed AO-based scheme distorts more slowly with increasing UE distance, compared to that of the benchmark schemes.

Fig. \ref{fig:Rician factor} shows the impact of the Rician factor on the achievable data rate. To compare beamforming design schemes without the influence of the BT process, we uniformly adjust the Rician factor of all inter-array channels while keeping the Rician factor of $\mathbf{h}^H$ fixed at 20 dB. The proposed AO-based scheme outperforms the benchmark schemes across different Rician factors, showing its effectiveness.
Moreover, as the Rician factor increases, the achievable rates for all schemes increase and then plateau. The increase is because only the inter-array LoS channels are used for the beamforming design, while the random NLoS components are not exploited. The plateau is because when the Rician factor is sufficiently large, the channel gradually converges with the LoS component.

\section{Conclusion}
In this letter, we have investigated the beamforming design to enhance the performance of multi-IRS-assisted wireless beam routing systems, taking into account practical near-field channel modeling. Under a beam-training-based communication protocol, the problem was formulated to maximize the received power by jointly optimizing the beamforming of the BS and intermediate IRSs. To address the formulated problem, a reformulation and an efficient AO-based algorithm were proposed. Numerical results have demonstrated that the power scaling derived under the far-field assumption no longer holds under near-field conditions. Furthermore, compared to the benchmark schemes, the proposed AO-based scheme obtains superior system performance by achieving higher received signal power at the UE.


\def\baselinestretch{1}
\bibliographystyle{IEEEbib}
\bibliography{IEEErefs}
\end{document}